\documentclass[aps,
twocolumn,prl,preprintnumbers,nofootinbib]{revtex4}

\usepackage{bbm}
\usepackage{bm}

\newcommand\lsim{\mathrel{\rlap{\lower4pt\hbox{\hskip1pt$\sim$}}
    \raise1pt\hbox{$<$}}}
\newcommand\gsim{\mathrel{\rlap{\lower4pt\hbox{\hskip1pt$\sim$}}
    \raise1pt\hbox{$>$}}}

\begin{document}

\title{Seesaw at LHC}

\author{Borut Bajc}
\email{borut.bajc@ijs.si}
\affiliation{J. Stefan Institute, 1001 Ljubljana, Slovenia}
\author{Goran Senjanovi\' c}
\email{goran@ictp.it}
\affiliation{International Centre for Theoretical Physics,
34100 Trieste, Italy }

\begin{abstract}
We study the implementation of the type III seesaw in the 
ordinary nonsupersymmetric SU(5) grand unified theory. This allows for 
an alternative definition of the minimal SU(5) model, with the 
inclusion of the adjoint fermionic multiplet. The main 
prediction of the theory is the light fermionic 
SU(2) triplet with mass at the electroweak scale.
 Due to their gauge couplings, 
these triplets can be  produced pair-wise via Drell-Yan, and due to the Majorana
nature of the neutral component their decays leave a clear signature of same sign
di-leptons and four jets. This allows for their possible discovery at LHC and provides 
an example of directly measurable seesaw parameters.

\end{abstract}


\maketitle

{\it A. Introduction}. \hspace{0.5cm}
We know today that neutrinos are massive (at least two of them). 
This implies that the minimal standard model cannot be the whole 
story. If one is not to change its low energy structure, one is 
led to a higher dimensional operator \cite{Weinberg:1979sa}
simbolically

\begin{equation}
Y_{ij}\frac{l_il_jHH}{M}\;,
\end{equation}

\noindent
where $l_i$ is the usual lefthanded leptonic doublet 
and $H$ the Higgs doublet of the SM. 
Demanding perturbativity, i.e. $Y_{ij}\lsim 1$ 
implies $M\lsim 10^{14}$ GeV, much below the Planck scale. 
In other words, gravity does not suffice and one must 
introduce new heavy states to be integrated out. This is 
called the seesaw mechanism. There are only three possible 
ways of implementing the seesaw: 

I) one introduces right-handed neutrinos (at least two) 
\cite{seesaw};

II) one utilizes a heavy SU(2) triplet with an appropriate 
hypercharge and a small vev \cite{Magg:1980ut}; 

III) one introduces heavy triplet fermions with zero 
hypercharge (at least two of them) \cite{Foot:1988aq}.

The first two possibilities, called type I and type II 
are being pursued daily, whereas the third one, called 
type III, has been very little discussed. The reason 
could be the necessity of having a number of such triplets, 
but even that may be weakened, if one accepts a combination 
of seesaw mechanisms. For example, a triplet and a 
singlet of fermions suffice to give two massive light 
neutrinos. Still, at first glance, it seems raher ad-hoc 
to use such a strange combination.

By itself, the seesaw mechanism sheds no light on 
neutrino mass, for it is equivalent to the effective 
operator written above. It is 
indispensable to have a theory beyond the standard model 
that predicts at least the scale $M$, if not the 
couplings themselves. The natural framework for such a 
theory is grand unification and the minimal grand unified 
group, as well known, is based on SU(5). Suppose that one 
wants to study the minimal such theory without introducing 
supersymmetry, i.e. the original theory \cite{Georgi:1974sy} 
with $24_H$ and $5_H$ and the three generations of 
$10_F$ and $\overline{5}_F$. This theory is ruled out 
since the couplings of the standard model do not unify 
and furthermore neutrinos are massless. Adding I) righthanded 
neutrinos does not help, unification still fails. When 
defining the minimal nonsupersymmetric SU(5) one normally 
resorts thus to the case II), i.e. one adds the $15_H$ 
dimensional Higgs. This has been studied recently at length 
\cite{Dorsner:2005fq}. 

The third possibility was not studied at all and this is 
the scope of our work. It amounts to adding a new set of 
fermions, $24_F$, and can be considered as an alternative 
minimal nonsupersymmetric SU(5) theory. 
This cures both the unification problem and 
accounts for a realistic neutrino spectrum. The reason 
for the latter is that $24_F$ contains both triplet and 
singlet fermions, and thus utilizing type III seesaw 
gives also type I as a bonus. 

Although the theory will require substantial fine-tuning, 
it turns out to be remarkably predictive. The combination 
of proton decay and unification constraints predicts the 
mass of the triplet fermion in $24_F$ and the mass of 
the triplet scalar in $24_H$ below TeV, likely to be 
found at LHC. This is the main and the most interesting 
prediction of the theory. The stability of the proton 
prefers these particles to lie as close as possible to $M_Z$. 

The masses of the other particles are also restricted. The 
colour octets are some $3$ to $6$ orders of magnitude 
heavier than the triplets, while the fermionic leptoquarks 
turn out to lie at the intermediate scale $10^{11-13}$ GeV. 

In short, this theory provides an interesting example of seesaw 
particles predicted to be detectable at LHC and their Yukawa 
couplings directly accessible. 

\vspace{0.2cm}

{\it B. The model}. \hspace{0.5cm} The minimal 
implementation of the type III seesaw in 
nonsupersymmetric SU(5) requires a fermionic adjoint 
$24_F$ in addition to the usual field content $24_H$, $5_H$ 
and three generations of fermionic $10_F$ and $\overline{5}_F$. 
The consistency of the charged fermion masses requires 
higher dimensional operators in the usual Yukawa sector 
\cite{Ellis:1979fg}. One must add the new Yukawa interactions 

\begin{eqnarray}
{\cal L}_{Y\nu}&=&
y_0^i
\left(\bar 5_F^i\right)
\left(24_F\right)
5_H+\frac{1}{M_{Pl}}
\left(\bar 5_F^i\right)
\left[
y_1^i24_F24_H\right.\nonumber\\
&+&\left.
y_2^i24_H24_F+
y_3^iTr\left(24_F24_H\right)
\right]
5_H\;.
\end{eqnarray}

After the SU(5) breaking (for later use 
$\langle 24_H\rangle =diag\left(2,2,2,-3,-3\right)v/\sqrt{30}$)
one obtains the following physical 
relevant Yukawa interactions for neutrino with the triplet 
$\sigma_3^F\equiv\overrightarrow{\sigma}_3^F\overrightarrow{\tau}$ 
(type III) and singlet $\sigma_0^F$ (type I) fermions:

\begin{equation}
\label{lynu}
{\cal L}_{Y\nu}=L_i\left(
y_\nu^{(3)i}\sigma_3^F+
y_\nu^{(0)i}\sigma_0^F\right)H\;,
\end{equation}

\noindent
where $y_\nu^{(3)i}$, $y_\nu^{(0)i}$ are two different 
linear combinations of $y_0^i$ and $y_a^i v/M_{Pl}$ 
($a =1,2,3$). 
It is clear from the above formula that besides the new 
appearence of the triplet fermion, the singlet fermion 
in $24_F$ acts precisely as the righthanded neutrino; 
it should not come out as a surprise, as it has the right 
SM quantum numbers. 

Even before we discuss the physical consequences in detail, 
one important prediction emerges: only two light neutrinos 
 get mass, while the third one  remains massless. 

In order to discuss the masses of the new fermions, we need 
the new Yukawa couplings between $24_F$ and $24_H$

\begin{eqnarray}
\label{lf}
{\cal L}_{F}&=&m_FTr\left(24_F^2\right)+
\lambda_FTr\left(24_F^224_H\right)\\
&+&\frac{1}{M_{Pl}}\left[
a_1Tr\left(24_F^2\right)Tr\left(24_H^2\right)
+a_2\left(Tr\left(24_F24_H\right)\right)^2
\right.\nonumber\\
&+&\left.
a_3Tr\left(24_F^224_H^2\right)+
a_4Tr\left(24_F24_H24_F24_H\right)\right]\;,\nonumber
\end{eqnarray}

\noindent
where we include the higher dimensional terms for the sake 
of consistency. The masses of the new fermions are 

\begin{eqnarray}
\label{m0}
m_0^F&=&m_F-\frac{\lambda_Fv}{\sqrt{30}}+
\frac{v^2}{M_{Pl}}\left[a_1+a_2+
\frac{7}{30}\left(a_3+a_4\right)\right]\;,\\
\label{m3}
m_3^F&=&m_F-\frac{3\lambda_Fv}{\sqrt{30}}+
\frac{v^2}{M_{Pl}}\left[a_1+
\frac{3}{10}\left(a_3+a_4\right)\right]\;,\\
\label{m8}
m_8^F&=&m_F+\frac{2\lambda_Fv}{\sqrt{30}}+
\frac{v^2}{M_{Pl}}\left[a_1+
\frac{2}{15}\left(a_3+a_4\right)\right]\;,\\
\label{m32}
m_{(3,2)}^F&=&m_F-\frac{\lambda_Fv}{2\sqrt{30}}+
\frac{v^2}{M_{Pl}}\left[a_1+
\frac{\left(13a_3-12a_4\right)}{60}\right]\;.
\end{eqnarray}

Next we turn to the bosonic sector of the theory. We will 
need the potential for the heavy field $24_H$

\begin{eqnarray}
\label{v24h}
V_{24_H}&=&
m_{24}^2Tr\left(24_H^2\right)+
\mu_{24}Tr\left(24_H^3\right)\\
&+&
\lambda_{24}^{(1)}Tr\left(24_H^4\right)+
\lambda_{24}^{(2)}\left(Tr\left(24_H^2\right)\right)^2
\nonumber\;,
\end{eqnarray}

\noindent
and its interaction with the light fields

\begin{eqnarray}
\label{v5h}
V_{5_H}&=&m_H^25_H^\dagger 5_H+
\lambda_H\left(5_H^\dagger 5_H\right)^2
+\mu_H5_H^\dagger 24_H5_H\nonumber\\
&+&
\alpha\left(5_H^\dagger 5_H\right)Tr\left(24_H^2\right)+
\beta5_H^\dagger 24_H^25_H\;.
\end{eqnarray}

It is a straightforward exercise to show that the 
masses of the bosonic triplet and octet are arbitrary and 
that one can perform the doublet-triplet splitting through 
the usual fine-tuning. 

We are now fully armed to study the constraints on 
the particle spectrum by performing the renormalization 
group analysis. 

\vspace{0.2cm}

{\it C. Proton decay and unification constraints.} \hspace{0.5cm} 
Before getting lost in the numerics, it is useful to 
recall the failure of the SM unification \cite{Langacker:1991an}. 
The weak and 
strong couplings actually unify at the scale around 
$10^{16}$ GeV, just as in the supersymmetric version of 
the theory. This is ideal for the proton decay point of view, 
but the trouble is that the U(1) coupling hits the weak coupling too 
soon, at the scale of about $10^{12-13}$ GeV. This indicates 
that the weak triplets are expected to be light in order 
to slow down the decrease of the weak coupling. It is easy 
to see that the fermionic leptoquark makes things worse 
and, as we show carefully below, they should be as heavy 
as possible. However splitting its mass from the triplet and 
the octet fermion masses require the inclusion of higher 
dimensional terms, which in turn gives an upper bound to the 
mass of the leptoquark 

\begin{equation}
m_{(3,2)}^F\lsim\frac{M_{GUT}^2}{M_{Pl}}\;.
\end{equation}

For the sake of illustration we present first the one-loop 
analysis. The renormalization group equations at 
this level are

\begin{eqnarray}
\label{alfa1}
&&2\pi\left(\alpha_1^{-1}(M_Z)-\alpha_U^{-1}\right)=
\frac{41}{10}\ln\frac{M_{GUT}}{M_Z}\\
&&+\frac{10}{3}\ln\frac{M_{GUT}}{m_{(3,2)}^F}
+\frac{1}{15}\ln\frac{M_{GUT}}{m_T}\;,\nonumber\\
&&2\pi\left(\alpha_2^{-1}(M_Z)-\alpha_U^{-1}\right)=
-\frac{3}{2}\ln\frac{M_{GUT}}{M_Z}\\
&&-\frac{4}{3}\ln\frac{m_3^F}{M_Z}
-\frac{1}{3}\ln\frac{m_3^B}{M_Z}
+2\ln\frac{M_{GUT}}{m_{(3,2)}^F}\;,\nonumber\\
\label{alfa3}
&&2\pi\left(\alpha_3^{-1}(M_Z)-\alpha_U^{-1}\right)=
-\frac{9}{2}\ln{\frac{M_{GUT}}{M_Z}}\\
&&-2\ln{\frac{m_8^F}{M_Z}}
-\frac{1}{2}\ln{\frac{m_8^B}{M_Z}}
+\frac{4}{3}\ln\frac{M_{GUT}}{m_{(3,2)}^F}
+\frac{1}{6}\ln{\frac{M_{GUT}}{m_T}}\;,\nonumber
\end{eqnarray}

\noindent
where $m_3^{F,B}$, $m_8^{F,B}$, $m_{(3,2)}^F$ and $m_T$ are 
the masses of weak triplets, colour octets, (only fermionic) 
leptoquarks and (only bosonic) colour triplets respectively. 

From the above a straightforward computation gives 

\begin{eqnarray}
\label{a12}
&&\exp{\left[30\pi\left(\alpha_1^{-1}-\alpha_2^{-1}\right)
\left(M_Z\right)\right]}=\\
&&\left(\frac{M_{GUT}}{M_Z}\right)^{84}
\left(\frac{\left(m_3^F\right)^4
m_3^B}{M_Z^5}\right)^{5}
\left(\frac{M_{GUT}}{m_{(3,2)}^F}\right)^{20}
\left(\frac{M_{GUT}}{m_T}\right)\nonumber\\
\label{a13}
&&\exp{\left[20\pi\left(\alpha_1^{-1}-\alpha_3^{-1}\right)
\left(M_Z\right)\right]}=\\
&&\left(\frac{M_{GUT}}{M_Z}\right)^{86}
\left(\frac{\left(m_8^F\right)^4
m_8^B}{M_Z^5}\right)^{5}
\left(\frac{M_{GUT}}{m_{(3,2)}^F}\right)^{20}
\left(\frac{M_{GUT}}{m_T}\right)^{-1}\nonumber
\end{eqnarray}

\noindent
where we still keep all the masses generic, including 
the one of the leptoquark. As we argued before, its mass 
must be at most of order $M_{GUT}^2/M_{Pl}$, which simplifies the 
analysis. From the well known problem in the standard 
model of the low meeting scale of $\alpha_1$ and $\alpha_2$, 
it is clear that the SU(2) triplet should be as light as 
possible and the colour triplet as heavy as possible. In 
order to illustrate the point, take $m_3^F=m_3^B=M_Z$ and 
$m_T=M_{GUT}$. This gives ($\alpha_1^{-1}(M_Z)=59$,
$\alpha_2^{-1}(M_Z)=29.57$, $\alpha_3^{-1}(M_Z)=8.55$) 
$M_{GUT}\approx 10^{15.5}$ GeV. Increasing the 
triplet masses $m_3^{F,B}$ reduces $M_{GUT}$ dangerously, making 
at the same time proton decay too fast and higher dimensional 
operators (needed to correct the second generation charged 
fermion masses) too small. 

The two loop effects \cite{Bajc:2007zf} relax this somewhat 
and for the above example of the GUT scale the triplet 
mass increases to about $500$ GeV. 
Even if one allows $M_{GUT}$ as low as $10^{15}$ GeV, one 
gets the triplet mass about few TeV. In 
this extreme case this particle would not be produced at LHC, 
but would make leptogenesis easier to function. We should 
stress though that one is really stretching the parameters 
in order to avoid this triplet be discovered at LHC. 

We can safely conclude that the SU(2) triplets, 
especially the fermionic one responsible for the type III 
seesaw, should lie close to $M_Z$ and possibly be detectable 
at LHC. This is the main result of our work. Simultaneously 
proton lifetime is predicted to be close to the experimental 
limit, since the GUT scale must lie below $10^{16}$ GeV. 
This makes a strong case for the new generation of proton 
decay experiments. 

From eq. (\ref{a13}) one finds the fermion colour octet mass 
in the range $10^5-10^8$ GeV, beyond experimental reach. The 
bosonic equivalent is actually not constrained by RGE at all 
and can be as light as $M_Z$. 
The solution we described here reminds the so called split 
supersymmetry \cite{Arkani-Hamed:2004fb} in the limit of very large 
higgsino masses. Due to their absence here the colour octet (the gluino 
in split supersymmetry) is much heavier that the weak triplet 
(the wino in split susy). 

\vspace{0.2cm}

{\it D. Phenomenological implications.} \hspace{0.5cm} 
The simplicity of the theory is reflected in the 
neutrino sector too. As we remarked, one neutrino 
is massless. This is true up to possible effects of 
gravity \cite{Akhmedov:1992hh}, but gravity can only 
give a mass of about $10^{-5}-10^{-6}$ eV, effectively 
zero for all practical purposes. 
The six complex parameters in (\ref{lynu}) 
($y_\nu^{(3)i}$, $y_\nu^{(0)i}$) become only nine real parameters 
after the redefinition of the leptonic phases. The 
model is thus similar to an often imagined situation 
of two righthanded neutrinos, only here it is predicted 
by the structure of the theory. 

Since the triplet $\sigma_3^F$ is at the weak scale, the 
couplings $y_\nu^{(3)i}$  are generically  of the order of 
$10^{-6}-10^{-7}$ (barring accidental cancellations) , whereas 
the couplings $y_\nu^{(0)i}$ depend on the mass of the singlet 
$\sigma_0^F$. This mass cannot be determined by the 
unification constraints,  because $\sigma_0^F$ 
is a SM gauge singlet. In any case, since one of the masses 
vanishes, the spectrum of light neutrinos corresponds either to 
the normal or inverse hierarchy. 

The most interesting predictions of the theory regards 
LHC. The fact that seesaw is achieved through a triplet 
has a remarkable impact. Since its mass is close to $M_Z$, 
its Yukawa couplings are very small and thus if it were 
a standard model singlet, it would be basically invisible. 
However, as an SU(2) triplet, it can be easily produced 
(if $m_3\lsim 500$ GeV \cite{denegri}) 
through the gauge interactions, and in this sense it 
behaves very much as a wino without higgsinos. 
These leptons would be produced in pairs 
through a Drell-Yan process. The production cross section for 
the sum of all three possible final states, $T^+ T^-$, $T^+ T^0$ 
and $T^- T^0$, can be read from Fig.2 of ref. \cite{Cheung:2005ba}: 
it is approximately 20 pb for 100 GeV triplet mass, and around 
50 fb for 500 GeV triplets. The triplets then decay through the 
same Yukawa couplings 
(\ref{lynu}) that enter into the seesaw. More precisely, after the 
SU(2) breaking the heavy triplet mixes with leptons and thus its 
main decays become 

\begin{eqnarray}
\left(\sigma_3^F\right)^-&\to &Z l^-\;,\;
W^-\nu\;\\
\left(\sigma_3^F\right)^0&\to &W^+l^-\;,\;Z\nu\;.
\end{eqnarray}

One can estimate 

\begin{equation}
\Gamma\left(\sigma_3^F\right)\approx |y_\nu^{(3)}|^2m_3^F\;,
\end{equation}

\noindent
which gives $\tau\left(\sigma_3^F\right)\approx 10^{-13}-10^{-16}$ sec. 
This leaves a clear signature at LHC, providing an important 
example of the seesaw mechanism being testable at TeV energies. 
The clearest signature is 
the three charged lepton decay of the charged triplet, but it 
has only a $3\%$ branching ratio. A more promising situation is 
the decay into two jets with heavy gauge boson invariant mass 
plus a charged lepton: this happens in  
approximately 23\% of all decays. 
The main point here is that the neutral component of the triplet decays
as often into a charged lepton as into an antilepton due to its Majorana
nature (just like right handed neutrinos).

The signatures in this case would be two same charge leptons plus two pairs 
of jets having the W or Z mass and peaks in the lepton-dijet mass. 
From the above estimates the cross section for such events is 
around 2pb (5fb) for 100 (500) GeV triplet mass. 
 Such signatures were suggested originally in 
the case of the type I seesaw in L-R symmetric theories \cite{Keung:1983uu},
but are quite generic of the seesaw mechanism. The only difference in the type I case is
that the dileptons are accompanied by two jets instead of four for the type III.

The colour octet fermions and bosons must decay before nucleosynthesis. 
It is easy to see that the bosonic octet decays through $1/M_{Pl}$ 
Yukawa couplings, which sets a limit $m_8^B\gsim 10^5$ GeV. If the 
fermionic octet is heavier than the bosonic one and the fermionic 
singlet together, then it can decay into them 
through the couplings in (\ref{lf}). If the opposite is true, the 
fermionic octet can decay through the exchange of the heavy colour 
triplet in $5_H$, which requires $m_T\lsim 10^{13}$ GeV. This would 
be yet another hope for an observable proton decay in the future. 

Although somewhat less firmly, the theory also predicts 
a light scalar triplet $\sigma_3^B$ from $24_H$. If stable, this 
would provide a classical example of an ideal dark matter 
candidate (wimp). Can it be stable? The answer is no due to 
the unavoidable presence of higher dimensional operators that 
correct the bad SU(5) fermion mass relations \cite{Bajc:2007zf}.

\vspace{0.2cm}

{\it E. Summary and outlook.} \hspace{0.5cm} 
In this letter we have constructed the minimal predictive 
SU(5) theory. It is based on the addition of an adjoint 
fermionic multiplet to the already existing bosonic adjoint and 
fundamental Higgses. Through the existence of the standard model fermion 
singlet and weak triplet, one obtains a combination of the type I 
and type III seesaw and thus one massless neutrino. 
The scale is too low for thermal 
leptogenesis \cite{Fukugita:1986hr} to work 
(for a generic discussion of leptogenesis with type III 
seesaw see \cite{Hambye:2003rt}) unless the singlet and 
triplet fermions are almost degenerate (resonant leptogenesis)
as explicitly shown for the case of right-handed neutrinos in 
\cite{Pilaftsis:2003gt}. 

The crucial 
prediction of the theory are the light weak fermionic and bosonic 
SU(2) triplets with masses around $M_Z$. 

Probably the most exciting aspect of this theory is that 
the decays of possibly observable seesaw particles will 
probe directly the Yukawa Dirac couplings of neutrinos. 
Thus the neutrino masses 
are correlated with observable phenomena at the TeV energies. 
Last but not least, this is simultaneously tied to the prediction 
of proton decay being observable in the next generation of 
experiments. We postpone the detailed phenomenological and 
cosmological analysis of all these issues for the future.

\vspace{0.3cm}

We are grateful to Abdesslam Arhrib, Gustaaf Brooijmans, 
Daniel Denegri and Francesco Vissani for 
very useful discussions, and Miha Nemev\v sek for computational 
and other help. The work of G.S. was supported in part 
by the European Commission under the RTN contract 
MRTN-CT-2004-503369; the work of B.B. was supported in 
part by the Slovenian Research Agency. B.B. thanks 
ICTP for hospitality during the course of this work.

\end{document}